\documentclass[sigconf, nonacm]{acmart}

\AtBeginDocument{%
  \providecommand\BibTeX{{%
    \normalfont B\kern-0.5em{\scshape i\kern-0.25em b}\kern-0.8em\TeX}}}

\setcopyright{acmcopyright}
\copyrightyear{2018}
\acmYear{2018}
\acmDOI{10.1145/1122445.1122456}

\acmConference[Woodstock '18]{Woodstock '18: ACM Symposium on Neural Gaze Detection}{June 03--05, 2018}{Woodstock, NY}
\acmBooktitle{Woodstock '18: ACM Symposium on Neural Gaze Detection,
  June 03--05, 2018, Woodstock, NY}
\acmPrice{15.00}
\acmISBN{978-1-4503-9999-9/18/06}

\acmSubmissionID{1452}

\begin{document}

\title[Machine Learning for Connecting Rough Sleepers to Services]{A Recommendation and Risk Classification System for Connecting Rough Sleepers to Essential Outreach Services}

\author{Harrison Wilde}
\authornote{Marked authors contributed equally to this research and paper.}
\email{h.wilde@warwick.ac.uk}
\orcid{0000-0002-4391-8204}
\affiliation{%
  \institution{University of Warwick}
  \streetaddress{Mathematical Sciences Building, CV4 7AL}
}

\author{Lucia Lushi Chen}
\authornotemark[1]
\email{lushi.chen@ed.ac.uk}
\affiliation{%
  \institution{University of Edinburgh}
  \streetaddress{Informatics Forum, 10 Crichton St, Edinburgh EH8 9AB}
}

\author{Austin Nguyen}
\authornotemark[1]
\email{austin.to.nguyen@gmail.com}
\affiliation{%
  \institution{TripAdvisor}
  \streetaddress{}
}

\author{Zoe Kimpel}
\authornotemark[1]
\email{zoe.a.kimpel@gmail.com}
\affiliation{%
  \institution{Northwestern University, Masters in Data Science Graduate}
  \streetaddress{}
}

\author{Joshua Sidgwick}
\email{joshsidgwick@gmail.com}
\affiliation{%
  \institution{The Alan Turing Institute}
  \streetaddress{British Library, 96 Euston Road}
}

\author{Adolfo De Unanue}
\email{unanue@itam.mx}
\affiliation{%
  \institution{Instituto Tecnologico Autonomo de Mexico}
  \streetaddress{}
}

\author{Davide Veronese}
\email{dveronese@hks.harvard.edu}
\affiliation{%
  \institution{Harvard Kennedy School, Master in Public Policy Candidate}
  \streetaddress{79 John F. Kennedy Street, Cambridge, MA}
}

\author{Bilal Mateen}
\email{bilal.mateen@nhs.net}
\affiliation{
  \institution{The Alan Turing Institute}
  \streetaddress{British Library, 96 Euston Rd, London NW1 2DB}
}

\author{Rayid Ghani}
\email{rayid@cmu.edu}
\affiliation{%
  \institution{Carnegie Mellon University}
  \streetaddress{}
}

\author{Sebastian Vollmer}
\email{svollmer@turing.ac.uk}
\affiliation{
  \institution{The Alan Turing Institute}
  \streetaddress{British Library, 96 Euston Rd, London NW1 2DB}
}

\renewcommand{\shortauthors}{Wilde, Chen, Nguyen, Kimpel et al.}

\begin{abstract}

Rough sleeping is a chronic problem faced by some of the most disadvantaged people in modern society. This paper describes work carried out in partnership with Homeless Link, a UK-based charity, in developing a data-driven approach to assess the quality of incoming alerts from members of the public aimed at connecting people sleeping rough on the streets with outreach service providers. Alerts are prioritised based on the predicted likelihood of successfully connecting with the rough sleeper, helping to address capacity limitations and to quickly, effectively, and equitably process all of the alerts that they receive. Initial evaluation concludes that our approach increases the rate at which rough sleepers are found following a referral by at least 15\% based on labelled data, implying a greater overall increase when the alerts with unknown outcomes are considered, and suggesting the benefit in a trial taking place over a longer period to assess the models in practice. The discussion and modelling process is done with careful considerations of ethics, transparency and explainability due to the sensitive nature of the data in this context and the vulnerability of the people that are affected.

\end{abstract}



\maketitle

\section{INTRODUCTION}

Homelessness and rough sleeping comprise a pressing and worsening global issue that negatively affects a population through a host of societal and health-related pressures spanning poverty, illness and abuse \cite{fetzer2019}. The United Nations Human Settlements Program estimates that 1.1 billion people are living in inadequate housing, and the available data suggests that more than 100 million people have no housing at all \cite{igh2019}. Homelessness affects people in every region of the world, developed and developing, and in the absence of government-level coordinated action it is likely to continue growing \cite{ortiz-ospina_homelessness_2019}.

Rough sleeping is \href{https://www.gov.uk/guidance/homelessness-data-notes-and-definitions}{defined by the UK government} as a category of homelessness referring to the act of sleeping in the open air or other places not designed for human habitation; carried out by people who do not have access to permanent, consistent shelter \cite{govdefn}. A rough sleeper is vulnerable even relative to the homeless population as a whole; they are more likely to experience violence, health-related issues, sexual exploitation, and substance abuse \cite{meinbresse2014}. Female rough sleepers are especially disadvantaged as they tend to be younger in age, require more mental health support than men, and are more likely to be victims of domestic violence. Because of this, female rough sleepers are anecdotally known to hide themselves for safety reasons, something that is shown to be true in Homeless Link's data. This behaviour decreases the known number of female rough sleepers in the UK and means it is often harder for them to find support \cite{pleace_women_2018}. On any given night in England, there are an estimated 4,700 people sleeping rough on the streets; this represents a 169\% increase in the rough sleeper population from 2010 to 2019 \cite{rsstats2019, ortiz-ospina_homelessness_2019}. Additionally, when censused, 60\% of those sleeping rough in London were new to the streets that night, further evidencing the transient and often spontaneous nature of homelessness.

Data-based approaches aimed at tackling and properly quantifying these issues are severely lacking; it has been shown that existing statistics disproportionately affect minorities and exhibit gender biases \cite{toro_toward_2007, toro_homelessness_2007}. \href{https://www.homeless.org.uk}{Homeless Link (HL) is a UK membership charity organisation}  working to end homelessness in England and Wales. One way HL seeks to achieve this mission is by operating  \href{https://www.streetlink.org.uk}{a platform called StreetLink} that serves as a conduit for communication between members of the public and over 300 local service providers (LSPs) spread across the UK. The platform allows for members of the public, as well as rough sleepers themselves, to submit an alert via phone, web, or mobile app regarding someone who is potentially sleeping rough.

The alerts are passed on to a team of volunteers at StreetLink who manually review them for quality before dispatching them as referrals to local service providers, who then attempt outreach and report on the outcome. This review process is based on the information contained in the alerts such as descriptions of the person and their location. Due to the resource constraints seen at the LSP level, only alerts that have sufficient information are turned into referrals. The review process is non-trivial and takes a significant amount of time, which can have a detrimental effect on the scale and speed of service that charities are able to provide. This is because much of the information in an alert will not necessarily be valid more than a day after it was submitted as rough sleepers move around or have their situations change at short notice. Moreover, the volume of alerts can become overwhelming during periods of extreme weather and at the peak of winter. The high volume of alerts means that they cannot effectively be processed despite the urgency of rough sleepers' needs, especially during these potentially more dangerous and difficult times.

The success of StreetLink depends on whether the local outreach teams can successfully find rough sleepers based on the information contained in referred alerts. According to StreetLink's historical data from 2012 to 2019, only 14\% of received alerts resulted in successfully connecting with a rough sleeper. Although this percentage has increased over time, it is still only around 20\% in 2019.


\section{CONTRIBUTIONS}

In collaboration with HL, our objective was to improve StreetLink by building a recommendation system based on the use of machine learning classifiers to automatically identify quality alerts as they arrive. In this case, a quality alert is one that includes sufficient information for LSPs to locate the rough sleeper in question. Without sufficient resources to review alerts in a timely manner, the quality of an alert decays as the information on the rough sleeper's location loses relevance over time. The majority of the alerts received by StreetLink originate in London, amounting to 65\% of all alerts; resources are especially constrained here making the problem all the more relevant.

If StreetLink volunteers and local service providers could focus their limited time, resources, and expertise on higher quality alerts, more rough sleepers could be connected with outreach services to alleviate their situations. Our solution ensures the quality of referrals as well as minimising any resource wastage by providing a clear ranking for StreetLink to follow in review. To achieve this, several machine learning classification algorithms were applied to two problems: identifying alerts that led to a referral being made, and of these referrals which ones resulted in a positive outcome. To summarise, our main contributions are:

\begin{enumerate}
    \item A novel approach to prioritising incoming alerts to maximise the chance that a rough sleeper is connected with, validating our models equitably using a bi-model approach to ensure that referrals and suggestions are made fairly across demographics.
    \item We identify a set of characteristics that could facilitate the establishment of a connection between rough sleepers and outreach teams. Previously untapped insights from HL's data could incite policy change and longer term positive impact.
\end{enumerate}

\subsection{Overview of the Existing Manual Approach}

StreetLink manages a phone line, website and mobile application which all feed alerts into a centralised system. The majority of these alerts are subsequently reviewed by volunteers for quality and the potential for duplication with existing alerts. Following discussions with volunteers, we can define their process as a search for the following three characteristics: 

\begin{enumerate}
    \item A location that is accessible for an outreach team and not near a known hotspot for street activity where regular outreach is done regardless.
    \item Evidence that the rough sleeper has bedded down or is likely to bed down in that location.
    \item Sufficiently helpful descriptions about the rough sleeper's appearance and location.
\end{enumerate}

These criteria are heuristic and based on the experiences of the outreach teams that StreetLink works with and represent a baseline decision making process for comparison. This process aims to ensure that the limited resources will go to individuals who have no choice but to sleep on the street instead of those who engage in street activities but already have a place in a shelter. In addition, StreetLink apply a naive rule-based algorithm to check for potential duplicated alerts prior to manual review (based on whether another alert was raised within 500 metres in the past week). StreetLink volunteers then confirm these duplicated cases during their review. 

\subsection{Data-Driven Approach}

\begin{figure*}
  \includegraphics[scale=0.25]{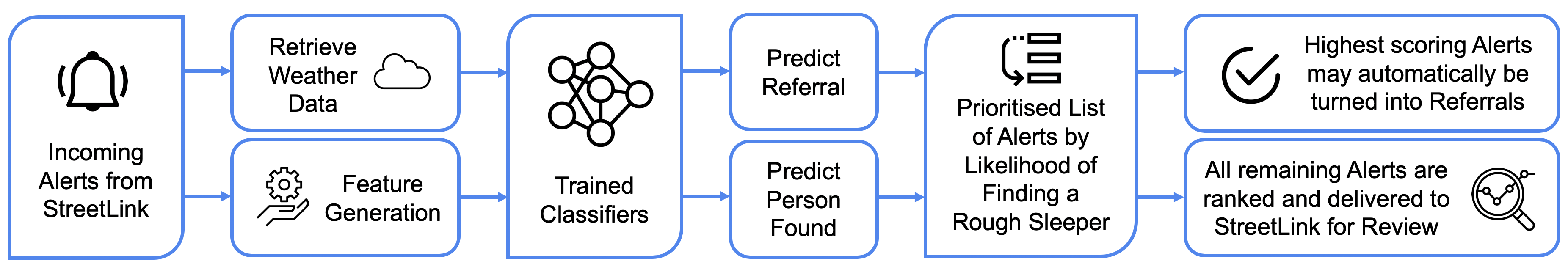}
  \caption{The proposed alert prioritisation process.}
  \label{fig:model}
\end{figure*}

The problem was formulated as: how can incoming alerts be best prioritised based on if they have sufficient information for outreach workers to connect with a rough sleeper, and whether the eventual outcome will be positive. Whilst duplicate alerts present a problem for volunteers by increasing their workload, we expect our models to utilise information in duplicate alerts co-operatively to gain a greater understanding of which features lead to positive outcomes. Additionally, it was necessary to build models to better understand the referral process carried out by StreetLink so as to ensure equity in the services delivered. These two components are classification problems with binary labels corresponding to whether a referral ends in a positive outcome (i.e. whether a rough sleeper will be found or not) and whether a referral was made, respectively.

Data provided by StreetLink was used to train a pair of models to be used in the review process shown in Fig~\ref{fig:model}:
\begin{enumerate}
    \item \textbf{Positive Outcome Model}: trained on binary labels indicating whether or not a person was found within a week following a referral; alerts that did not turn into referrals were therefore excluded from the training data for this model.
    \item \textbf{Referral Model}: trained on binary labels indicating whether or not an alert was turned into a referral by StreetLink.
\end{enumerate}

The Referral Model's purpose lies in emulating the current process of reviewing incoming alerts at StreetLink, and can be used alongside the Positive Outcome Model to highlight alerts that should have been turned into referrals as they were likely to have led on to positive outcomes. This system can be leveraged by StreetLink in the following way:
\begin{enumerate}
    \item Alerts with a high enough score from the Positive Outcome Model can automatically be sent to local service providers without manual review at the discretion of StreetLink. This frees up resources for volunteers to spend more time on reviewing and following up on the more nuanced and challenging alerts.
    \item All other incoming alerts should be ranked based on the Positive Outcome Model's scoring so that StreetLink volunteers review the most promising alerts first in order to best use the information that they contain by quickly turning these alerts into referrals.
\end{enumerate}

\begin{figure}
\includegraphics[width=80mm, scale = 1]{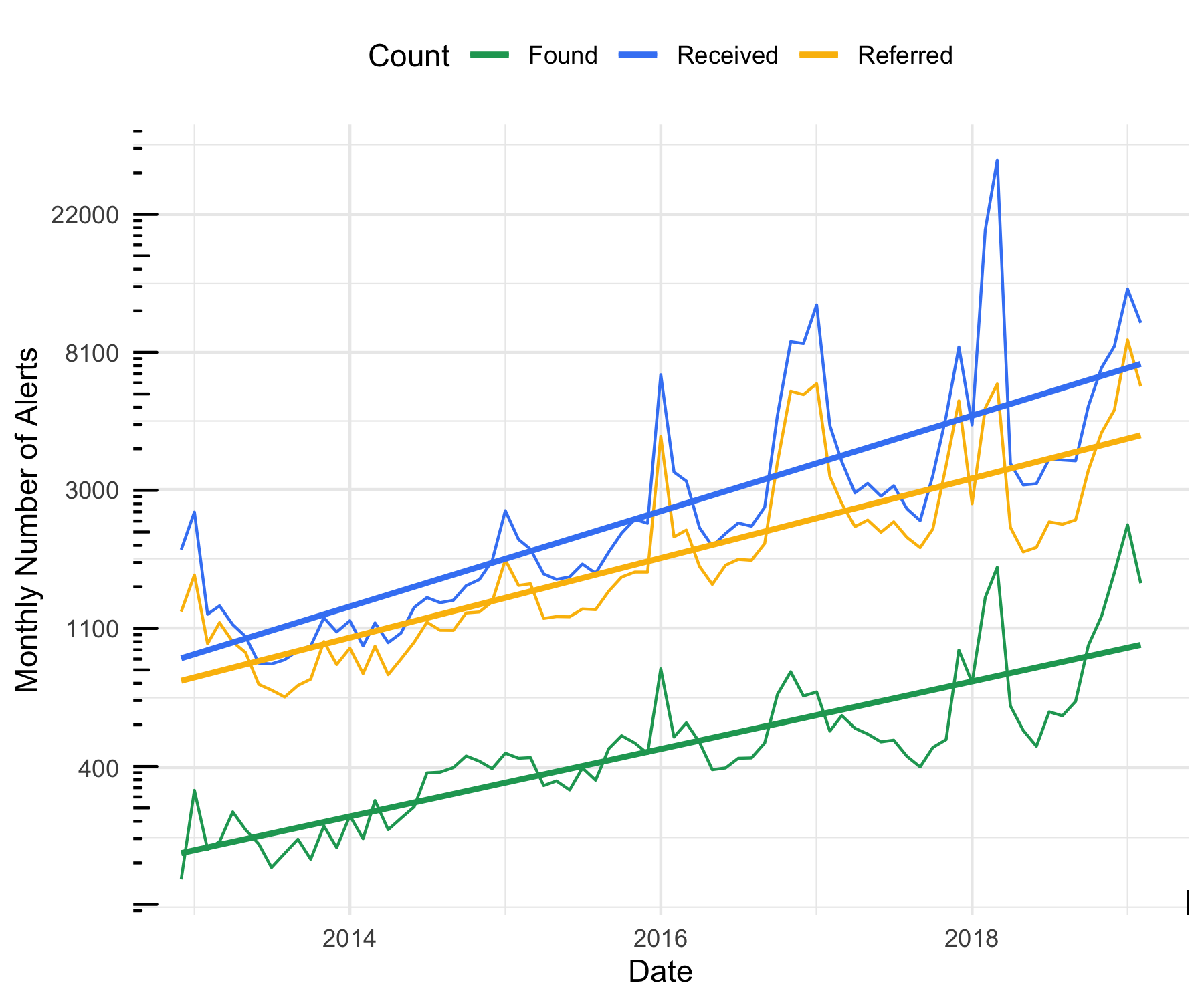}
\caption{StreetLink's platform is experiencing exponential growth, likely due to increased awareness of the problem and StreetLink itself. This increased level of demand has not been matched with an increased level of resources. The number of referrals made and people connected with follow a similar but slightly smaller exponential growth indicating the severity of the challenge that StreetLink faces.}
\label{fig:per}
\end{figure}

\section{Data}

\subsection{Primary Data}

Data was collected through StreetLink's alert reporting process from 2012 to 2019. The primary dataset contains 281,000 alerts, of which 167,000 are relevant to London and 114,000 are for non-London regions. 170,000 alerts were turned into referrals and of these 39,000 resulted in a positive outcome in which a person was found. Figure \ref{fig:per} shows the number of alerts, referrals and positive outcomes per month from 2012 to 2019. There is a strong seasonal effect on the volumes due to the public's increased awareness of rough sleepers during winter and the strain placed on service providers during this time. There was a significant operational change in December 2017 that altered the way in which StreetLink collected data; for the remainder of this paper we define our dataset to be the alerts from December 2017 to March 2019 (the end of the full dataset).

The raw data contains 43 fields that fall into the following five categories:
\begin{itemize}
    \item Demographics (age and gender)
    \item Outcomes and labels (signifying whether a person was found or not, or if a referral was created, as well as a number of other possibilities)
    \item Temporal information (time of alert creation and resolution, time a rough sleeper was seen at the location described)
    \item Location data (latitude and longitude provided by the user placing a pin on Google maps, full street address)
    \item Free text data (appearance description, location description, immediate concerns about the rough sleeper)
\end{itemize}

\subsection{External Data}

In addition to the data provided by StreetLink, we retrieved historical weather data for London (latitude: 51.50,  longitude: 0.1278) through \href{https://darksky.net/dev}{DarkSky's API}. The weather data ranges from 2012 to 2019, and includes hourly information about the temperature, wind speed and precipitation.

\subsection{Data Storage and Integration}

The data was stored in a PostgreSQL database in a Tier 3 Secure Environment hosted by The Alan Turing Institute, which helps prevent data loss and security breaches, see \cite{2019TuringSafehaven}. Multiple schema were created to store the data, features, and model results. 

\subsection{Data Preparation}

\paragraph{Cleaning the primary dataset} The historical data used throughout is an extract from StreetLink's Salesforce database. We renamed and formatted the variables for consistency, standardised text strings to lowercase, ensured empty values were recorded as NULLs, and standardised the format of temporal variables.

\paragraph{Labels} We created two levels of labels corresponding to each model discussed above. The first of which is a binary categorical variable capturing whether an alert was turned into a referral. The second label is also a binary categorical variable representing whether the outcome was positive, albeit one that can take NULL values in certain scenarios: where a referral was not made and the outcome is essentially unknown. The data originally contained 24 different outcome values that were mapped appropriately in order to define these binary labels. A number of alerts did not have clear outcomes, including alerts that were not handled by local authorities due to a lack of resources or that were still open following a referral; ee removed these alerts from the training dataset. Examples of some non-sensitive mappings are provided in Table~\ref{tab:label}.

\begin{table}
\caption{Examples of label mappings for the Referral and Positive Outcome Models}
\label{tab:label}
\begin{tabular}{lcc}
\toprule
Outcome                          & Referral Label & Pos. Outcome Label \\
\midrule
Person Found                     & Yes          & Yes               \\
LSP Did Not Respond             & Yes          & NULL              \\
Person Not Found                 & Yes          & No                \\
Not Enough Information           & No           & NULL        \\
\bottomrule
\end{tabular}
\end{table}

\subsection{Features}

270 features were created in total that can be split into five groups. All of the categorical variables representing demographic information were converted into dummy variables. Date-time features were created in 2 dimensions using sine and cosine functions to ensure consistent spacing between the last day of one month and the first day of the next, etc. Word counts of each free text field were also included.

\subsubsection{Spatio-Temporal Features}

\paragraph{Distances from Known Hotspots}

Known hotspots (places where StreetLink will not send alerts for as they know regular outreach is already carried out) were extracted using one of the outcomes in the data. Following this, a set of spatial variables was created to indicate if an alert was raised within $x$ meters of a known hotspot, $x\in X := \{50, 100, 250, 500, 1000, 5000, 10000\}$.

\paragraph{Alerts, Referrals, and People Found by Geographic Location, Time, and Source}

Similarly to the features described above, we generated spatio-temporal aggregate features to identify the number of alerts, referrals and positive outcomes within $x\in X$ metres, within the last $y\in Y$ days and for $z \in Z$ sources of alerts, $Y := \{7, 28, 60, 360\}$ and $Z := \{\text{Phone, Website, Mobile App}\}$. One example of this is the count of alerts received via phone within the last 28 days and within 100 metres of a given alert.

\paragraph{LSP-level Features}

Additionally, a set of features was created to represent the response statistics for different local service providers, including the average response time (based on the time between receipt of a referral and an outcome being reported), alert count, referral count, positive outcome count, and person found rates for each LSP. These features were again created over various $y\in Y$ temporal windows to capture trends in efficacy and activity at different resolutions.

\subsubsection{Other Features}

\paragraph{Weather}

The aforementioned weather data was aggregated by day to create the following features:
\begin{itemize}
    \item Temperature (maximum, minimum, average)
    \item Precipitation probability (maximum, minimum)
    \item Binary flag for the presence of any snow accumulation
    \item Wind (average speed, maximum gust)
\end{itemize}
Missing values were imputed by carrying forward the most recent observations in these cases.

\paragraph{Topic Features}

Alerts that provide more information on the location and activity of a rough sleeper are in general more useful to local service providers. Location and activity-related entities were extracted using pre-trained entity embeddings from the Python package SpaCy \cite{honnibal2015improved}. These entities were then grouped via an unsupervised learning technique --- Latent Dirichlet Allocation (LDA) (using the Gensim package in Python \cite{rehurek_lrec}). LDA is a generative statistical model to identify groups of topics in the free text fields. In this application, each field was viewed as containing a mixture of location and verb-related topics \cite{blei2003latent}. The algorithm estimates a score to represent the likelihood of an alert's fields belonging to a certain topic. These probability scores were then used as feature variables in the models.

LDA was carried out with various numbers of topics and it was found that the 10 topic solution was most representative of the extracted entities. The location topics encompassed entities including names of parks, hotels, train stations, streets, and so on. The same technique was used to extract 10 topics from the activity descriptions, hoping to separate alerts into those that described someone sleeping rough explicitly and those that described other street activity. These extracted activity-related topics included begging and sleeping, but many of the topics included the same activities. Therefore, we opted to also manually define two sets of topics to identify the activities of the rough sleepers: one containing sleep-related words ("tent", "duvet", etc.) and another including words related to begging behaviors ("beg", "small change", etc.). Word counts for all of these entities and LDA topics were used as features.

\section{MODELLING}

Due to the temporal dependencies present in the data, a month forward-chaining temporal cross validation approach \cite{roberts2017cross, varma_bias_2006} was adopted to ensure our model error estimates were robust across the entirety of the dataset. The alerts were split by month into folds of increasing length, beginning with the first month as training / validation data and the following month as test data for evaluation. Each subsequent fold's training data is defined as the concatenation of the previous fold's test and training data; it then uses the next month in sequence as test data, and so on. Model performance was then calculated as summary statistics of the performance across all folds. This idea generalises to varying periodicities but was carried out as described for a balance in robustness and model performance.

Additionally, there is a time span between an alert's creation and an outcome being provided. Therefore, for each train and test subset within a fold we needed to define a period within which an outcome would be accepted; removing other alerts that were open as of the date defining the endpoint of the training set. In our experiment we set this to be a week. This was necessary in order to avoid a clairvoyant model that could be trained on alerts that had outcomes which occurred during the period defined by its test set. As such, a week long buffer was maintained between the train and test sets as well as at the end of the test set, to allow for the final alerts in each subset to also have outcomes within our defined period of a week. 

A grid search of parameters was carried out for a series of classification algorithms, including ensemble models (Random Forest, Extra Trees), gradient boosting models (Adaptive Boosting), Decision Trees and dummy classifiers picking at random in a stratified manner consistent with the training data labels.

\section{EVALUATION}

\begin{table}
\caption{Metric Definitions}
\label{tab:metrics}
\begin{tabular}{p{0.18\linewidth}p{0.75\linewidth}} 
\toprule
Metric & Description  \\ [0.5ex] 
\midrule
Precision at \textit{\textbf{k}} & Total number of people found in the top \textit{\textbf{k}} alerts sorted by the model output scores and divided by the total number of people found or not found in the top \textit{\textbf{k}} alerts (excluding NULL labelled alerts) \\ 
\midrule
Found Rate at \textit{\textbf{k}} & Number of people found in the top \textit{\textbf{k}} alerts sorted by the model output scores and divided by \textit{\textbf{k}} (including NULL labelled alerts) \\
\midrule
Recall at \textit{\textbf{k}} & Number of people found in the top \textit{\textbf{k}} alerts sorted by the model output and divided by the total number of people found in that period's alerts \\
\bottomrule
\end{tabular}
\end{table}

Thousands of classification models were trained and evaluated for the two classification tasks. The models generate lists of alerts ranked according to their predicted likelihood of a positive outcome, or of a referral being made. Models can then be \textit{evaluated at varying \textbf{k}} by supposing that the model's top \textit{\textbf{k}} alerts by score are its suggested referrals. Then the metrics evaluated on these alerts can be compared to the real statistics from StreetLink (when \textit{\textbf{k}} = \textit{the number of referrals made by Homeless Link that fold / month}) and with other models. It is impossible to calculate true precision and recall for the Person Found Model due to the presence of NULL labels in the data where referrals were not made; we opt instead for the altered metrics defined in Table \ref{tab:metrics}. Since each model involves training a set of nested models through the aforementioned cross validation technique, the metric values reported in the results tables are the averaged values across all folds. Our objective is to choose a model which maximises all of the metrics, but with a particular focus on recall due to the implications of missing a positive outcome that StreetLink did not.

\begin{table}
\caption{Baseline Homeless Link Statistics by Fold / Month.}
\begin{tabular}{@{}lccc@{}}
\toprule
Fold / Month           & Positive Outcomes & Referrals & Found Rate \\ \midrule
January 2018   & 741               & 2707      & 0.2737                 \\
February 2018  & 1373              & 5421      & 0.2533                 \\
March 2018     & 1706              & 6442      & 0.2648                 \\
April 2018     & 625               & 2279      & 0.2742                 \\
May 2018       & 524               & 1909      & 0.2745                 \\
June 2018      & 467               & 1972      & 0.2368                 \\
July 2018      & 599               & 2373      & 0.2524                 \\
August 2018    & 582               & 2332      & 0.2500                 \\
September 2018 & 646               & 2407      & 0.2684                 \\
October 2018   & 970               & 3448      & 0.2813                 \\
November 2018  & 1199              & 4532      & 0.2646                 \\
December 2018  & 1639              & 5334      & 0.3073                 \\
January 2019   & 2323              & 8867      & 0.2620                 \\
February 2019   & 1521              & 6331      & 0.2402                 \\ \bottomrule
\end{tabular}
\label{tab:slmonthly}
\end{table}

\subsection{Defining the Baseline}

For the Referral Model, there is no meaningful baseline to compare against as it is impossible to beat StreetLink's human review process whilst using the labels defined by them. However, the number of referrals made by Homeless Link on a monthly basis can still provide some insight on how well our models emulate their referral process.

For the Positive Outcome Model, we can consider metrics at \textit{\textbf{k}} as defined in Table \ref{tab:metrics} and define two baselines:

\begin{enumerate}
    \item StreetLink's manual review process, where we can compare our models' found rates to that of StreetLink by observing the models at \textit{\textbf{k}}'s matching the monthly referrals by StreetLink shown in Table \ref{tab:slmonthly}. The ability to also look at our models' found rate and precision at lower values of \textit{\textbf{k}} for each month is compelling in justifying the partial automation of referral for the alerts that the models assign high scores to. However, it is difficult to formulate any rigorous comparisons with the baseline at lower \textit{\textbf{k}} due to the fact that StreetLink do not currently rank the referrals they send in a meaningful way.
    \item In order to compare precision and recall more clearly, a somewhat trivial baseline was formulated using a stratified dummy classifier that predicts based on the distribution of training labels. This baseline can be used for both referrals and alerts.
\end{enumerate}

\begin{table}
\caption{Results table for the best Positive Outcome Model. All values are averages across all of the temporal folds that the model was trained on.}
\label{tab:PO_result}
\begin{tabular}{@{}lcccc@{}}
\toprule
\textit{\textbf{k}}    & Precision & Recall & Found Rate & NULL Count \\ \midrule
50   & 0.7978         & 0.02679     & 0.5871          & 13            \\
100  & 0.7757         & 0.04932     & 0.5443          & 30            \\
150  & 0.7487         & 0.0708      & 0.5162          & 47            \\
200  & 0.7233         & 0.09118     & 0.4968          & 63            \\
300  & 0.7004         & 0.1283      & 0.4650          & 102           \\
400  & 0.6760         & 0.1591      & 0.4343          & 145           \\
500  & 0.6615         & 0.1867      & 0.4137          & 189             \\
750  & 0.6404         & 0.2560      & 0.3807          & 305           \\
1000 & 0.6209         & 0.3193      & 0.3606          & 419           \\
1500 & 0.5833         & 0.4406      & 0.3325          & 643           \\
2000 & 0.5548         & 0.5550      & 0.3129          & 870            \\
3000 & 0.5423         & 0.6830      & 0.3031          & 1309            \\
4000 & 0.5683         & 0.6785      & 0.3030          & 1847            \\
5000 & 0.5785         & 0.6615      & 0.3063          & 2308            \\
6000 & 0.5816         & 0.6668      & 0.2911          & 2950            \\
7000 & 0.5660         & 0.7373      & 0.2757          & 3531            \\ \bottomrule
\end{tabular}
\end{table}

\begin{table}
\caption{Results table for the best Referral Model. All values are averages across all of the temporal folds that the model was trained on.}
\label{tab:RM_result}
\begin{tabular}{@{}lcc@{}}
\toprule
\textit{\textbf{k}}     & Precision & Recall \\ \midrule
50    & 0.9457         & 0.01474     \\
100   & 0.9450         & 0.02947     \\
250   & 0.9406         & 0.07357     \\
500   & 0.9386         & 0.1467      \\
750   & 0.9381         & 0.2200      \\
1000  & 0.9348         & 0.2921      \\
1500  & 0.9204         & 0.4294      \\
2500  & 0.8688         & 0.6275      \\
5000  & 0.8115         & 0.6663      \\
7500  & 0.6975         & 0.7584      \\
10000 & 0.5664         & 0.8156      \\ \bottomrule
\end{tabular}
\end{table}

\begin{figure}
\includegraphics[width=85mm, scale = 1]{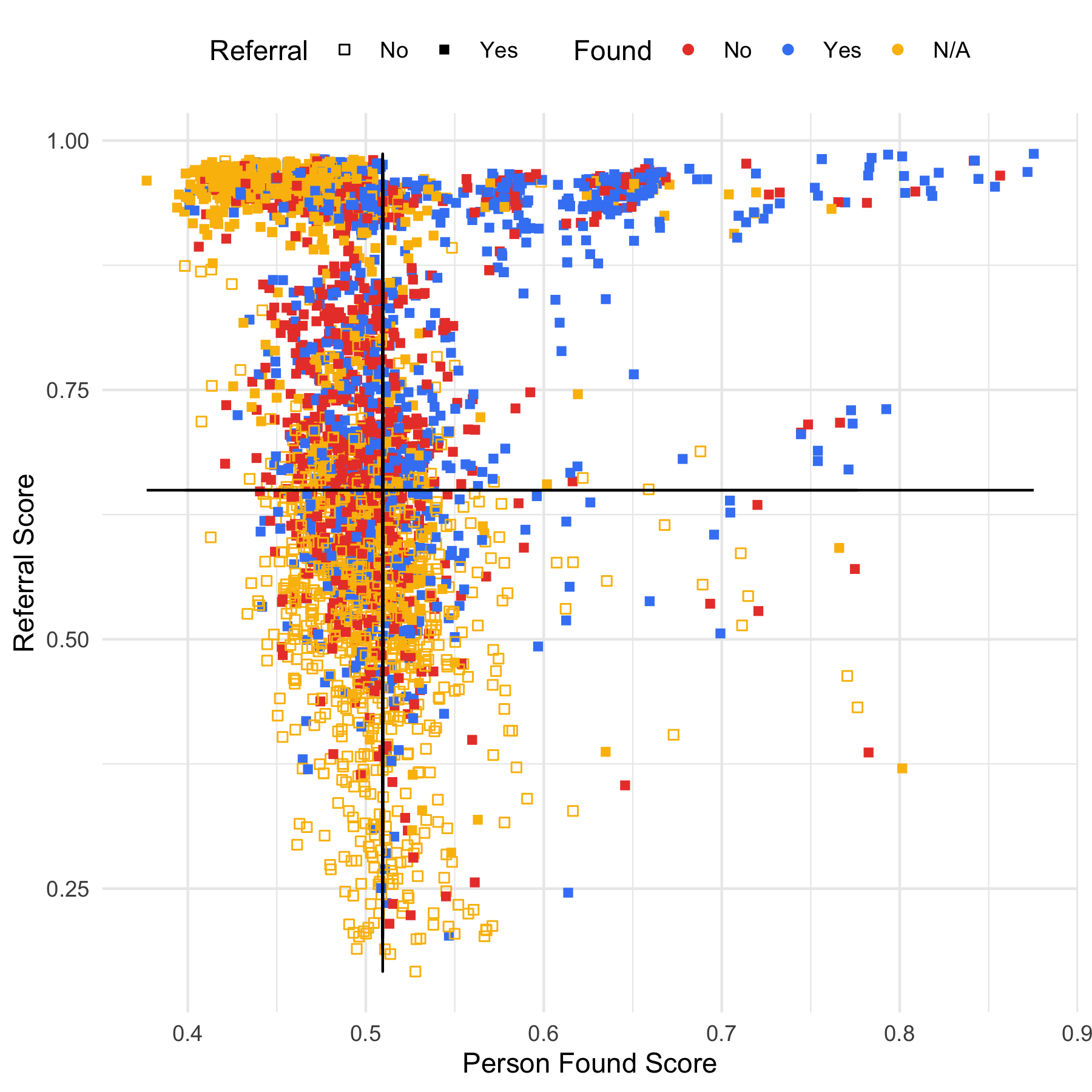}
\caption{Plot illustrating the distribution of scores for the chosen Positive Outcome and Referral models. This particular plot is for February of 2019.}
\label{fig:quadrant}
\end{figure}

\begin{figure*}
\includegraphics[width=175mm]{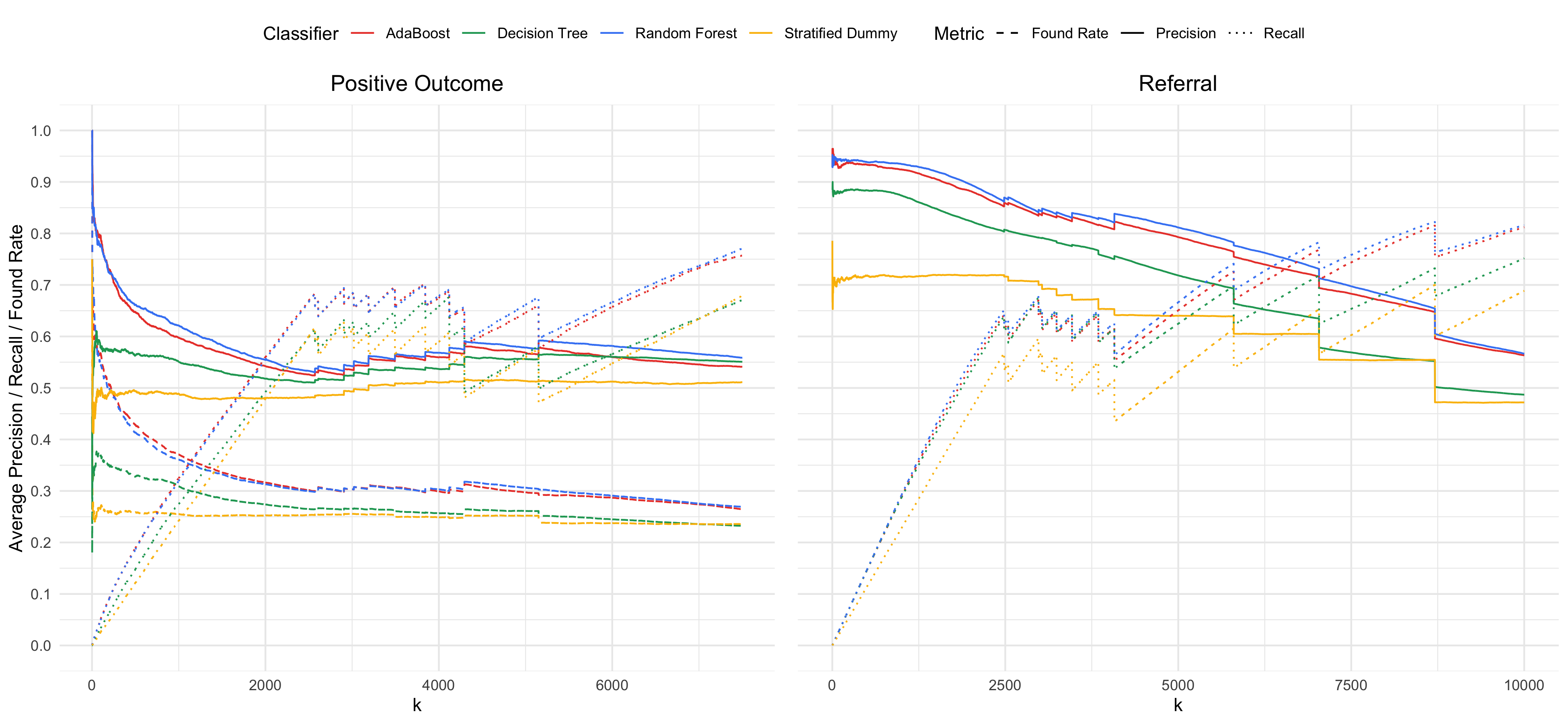}
\caption{The average precision, recall and found rate across all folds for the best examples of each classifier type. Points of seeming discontinuity arise due to the nature of our temporal nested cross validation; each fold and its corresponding test set spanning a month contains a different number of alerts.}
\label{fig:best}
\end{figure*}

\subsection{Model Comparisons and Choices}

The chosen Positive Outcome Model was a Random Forest Classifier with 10,000 trees and a maximum tree depth of 5 (see Table \ref{tab:PO_result}). The chosen Referral Model was a Random Forest Classifier with 10,000 trees and a maximum tree depth of 10 (see Table \ref{tab:RM_result}). To arrive at these choices we compared models on the averaged metrics defined in Table \ref{tab:metrics} and shown in Figure \ref{fig:best}, as well as through a number of other means described below.

Monthly statistics for StreetLink shown in Table \ref{tab:slmonthly} were used to compare found rates for the Positive Outcome Models and the real found rates across each fold. Here the Random Forest classifiers performed the best consistently, especially those with a large number of trees.

Figure \ref{fig:quadrant} is an example of a type of plot used for evaluation in which scores from each of the chosen models are plotted against each other and points are coloured according to the true Positive Outcome labels. The black lines are representative of the baseline in that there are a number of points to the right of the vertical line equal to the real number of people found in that month / fold. Similarly, there are an equal number of points above the horizontal line to the real number of referrals made in that month / fold. It can be seen that the majority of NULL-labelled points fall into the Positive Outcome score range of 0.45 to 0.5, whilst a lot of the alerts with a positive label are found in the top right quadrant of the graph. Interestingly, there are a significant number of alerts (approximately 20\%) within the bottom right quadrant that were not made into referrals per our Referral Model \textit{or} the true label, but that our Positive Outcome Model suggests would have led to positive outcomes with reasonable certainty. One of the initial objectives of this work was to increase StreetLink's efficiency so that they are able to process more referrals and potentially explore different types of referrals than they usually would; this quadrant would be a good place to start.

Further analysis by quadrant reveals that the top quadrants for Figure \ref{fig:quadrant} and similar graphs for each fold contain alerts with higher word counts in the free text fields than the average. The top right quadrant tends to include alerts that fulfil the criteria initially outlined in the Overview of the Existing Manual Approach section, as well as more alerts that have an unknown gender and age label. This is a different property to alerts that are missing these labels, as it explicitly suggests that whoever made the alert could not determine the rough sleeper's age or gender, possibly indicating that the person is sleeping and covered up. Moreover, there are a higher proportion of alerts corresponding to females in the bottom right quadrant than in any other, which further highlights the potential of this quadrant for exploration should Homeless Link have extra resources to spare in order to tackle the biases mentioned in the introduction. The chosen models all show promise in these areas and were otherwise minimal in the biases that they exhibit.

When comparing the two chosen models for each classification task, it can be seen that the Referral Model performs a lot better than the Positive Outcome Model at comparable \textit{\textbf{k}}. This is likely due to the fact that the referral process underlying the labels for the former is much simpler than the distribution defining whether an outcome of a referral will be positive or not. This is due to the complexity and number of factors that go into determining whether a person will be found when outreach is attempted.

Jaccard similarities were calculated for the label predictions made across different classifier types and configurations to assess whether different classifiers were better at predicting the outcomes of certain alerts. The results of this analysis were largely inconclusive with very few significant differences between lists of predictions; implying that some alerts are consistently more difficult to predict the outcomes of.

\subsection{Feature Importance}

\begin{figure}
\includegraphics[width=85mm, scale = 1]{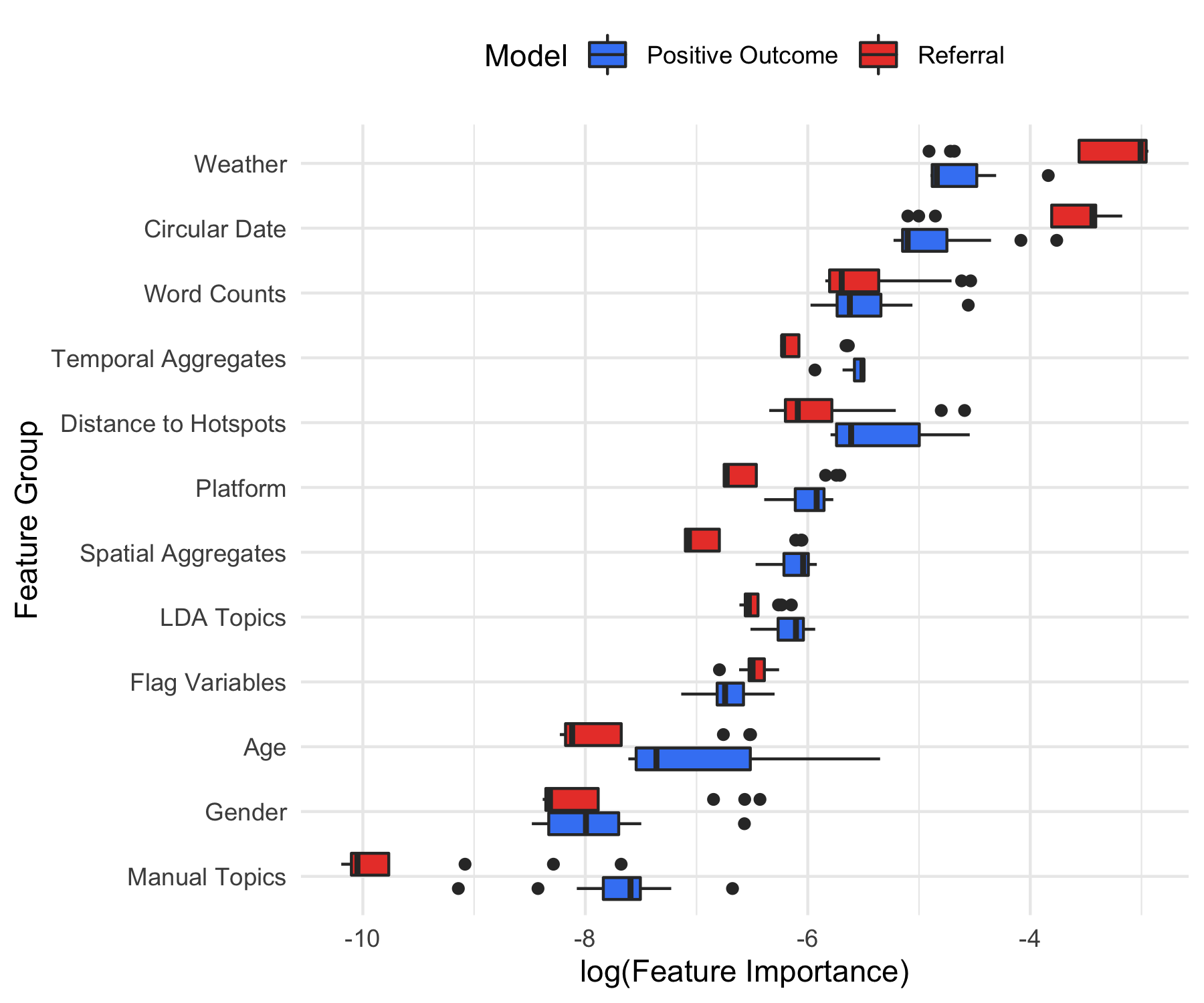}
\caption{Log-Scaled Feature Group Importances from each fold for the chosen Random Forest models. Temporal Aggregates includes all of the features generated by counting positive outcomes, referrals and total number of alerts received in varying time windows; Spatial Aggregates is similar but with varying proximities; LDA Topics include location and activity topics extracted from free text; Manual Topics are ones defined manually as important key words to count; Platform includes the features that indicate whether the alert originated from the web, mobile app or via a phone call.}
\label{fig:feature}
\end{figure}

One reason for the eventual model choices over the similarly performing AdaBoost classifiers is the ease with which their feature importances can be extracted. Figure \ref{fig:feature} shows the log-scaled feature importance scores for each feature group. Both of the top two features could feasibly represent seasonal effects and extreme weather, which often results in increased levels of demand and subsequently more outreach. Intuitively, word counts in the free text descriptions are likely to be indicative of the overall usefulness of an alert.

Different feature groups were included and excluded in various combinations to investigate the importances and interactions between features across repeated experiments. It was deemed justifiable to include all of the created feature groups in the training of the final models as many of the groups help to expose and illustrate potential biases in the recommendation system as well as carrying fairly intuitive real world interpretations.

\section{DISCUSSION}

This paper introduces a novel approach to assist StreetLink in connecting rough sleepers with local service provider outreach teams. Our Positive Outcome model generates a list of prioritised alerts that can be used by StreetLink staff to accelerate the existing manual review process and potentially augment it through automation. We also find that weather is the most essential characteristic in identifying the outcome of an alert; weather has the potential to be used as part of a forecasting system to help assess the need for rough sleeper outreach in anticipation of extreme weather across the country. 

The bi-model approach suggests that some of the alerts are currently unlikely to become a referral but have a high likelihood of having a positive outcome. These predictions are in conflict with the existing review process employed by StreetLink; we found that a majority of these alerts were not labelled as referrals in the manual review process. Part of the reason could be that some of these alerts were identified as duplicates; the existing system does not update the outcome of duplicated alerts. Furthermore, high quality alerts may not be sent as referrals during peak season due to an increased volume of alerts and associated resource constraints. Therefore, the manually assigned labels pose a significant challenge to the training and evaluation of our models due to the uncertainty present in a lot of the outcomes. The models will facilitate the collection of better data to begin filling these label gaps following deployment.

\subsection{Ethical Considerations}

The project was given a favourable opinion by the Alan Turing Institute's Ethics Advisory group. Potential disclosure concerns were mitigated through the use of the aforementioned Tier 3 secure computational environment. Moreover, all reported results conform to the UK's Office for National Statistics (ONS) disclosure risk mitigation guidance \cite{onsriskmanagement}. For example, aggregate values are not presented for any group smaller than 5 distinct individuals due to risk of re-identification of specific individuals.

We find that a large number of alerts regarding female rough sleepers are also characterised by this trend of a low referral score but a high positive outcome score. This reinforces the initial discussion on the trend of female rough sleepers being less likely to be connected with services as a result of hiding away from the public due to safety concerns. This amongst the other biases between the quadrants of Figure \ref{fig:quadrant} present opportunities for HL to examine whether they reach all demographics as equitably as possible.

Future work could focus on investigating whether rough sleepers within these demographics are willing to be contacted by outreach teams. If they also require help from local service providers, the existing process should be improved in order to reach people who do not feel safe to exist openly in public spaces. These biases in general pose ethical concerns about excluding certain groups of people from the current outreach support system.

Due to many of the alerts received by StreetLink containing information about the description and location of vulnerable people sleeping rough, ethical concerns were raised throughout the project and we discussed this issue as a team with Homeless Link and StreetLink volunteers. It is important to note that:
\begin{itemize}
    \item All alerts and outcome information was created with the consent of the alert author and the local service provider in the StreetLink application.
    \item Individuals contacted by local service providers have the ability to refuse any service or action at any time.
    \item To protect the sensitive information in the dataset, all of the data was hosted in a Tier 3 secure environment by The Alan Turing Institute with access limited to approved individuals and in line with data privacy regulations \cite{2019TuringSafehaven}.
    \item The primary goal of the partnership with HL was to positively impact the lives of rough sleepers in the UK by improving Homeless Link and StreetLink's process connecting people sleeping rough with local service providers.
    \item Concerns such as inappropriate use of the proposed models for harassment or tracking purposes of individual rough sleepers was discussed and is considered a strict violation. As such, use of these models is restricted to StreetLink staff and a small team of approved researchers and maintainers.
    \item Concerns were also raised that the models may learn any unknown biases in the existing human-centered process. For this reason, bias metrics were evaluated for the various demographics present in the data.
\end{itemize}

\subsection{Further Development}

The Positive Outcome Model's precision falls off beyond a  \textit{\textbf{k}} of 1000 to 2000. This shows a potential for improvement of this model by leveraging more complex NLP and spatio-temporal modelling approaches. Additionally, future work could pursue the issue of duplicates more explicitly by investigating whether unsupervised machine learning techniques can be used to cluster incoming alerts in spatial and temporal dimensions to give a better indication as to whether they might be duplicates with previous alerts: an issue which can only become increasingly critical as the use of StreetLink grows alongside public concern for the homeless.

Future work can also focus on engineering additional features from external data sources such as census and demographic data. As shown in our evaluation, the feature importances of weather related features ranked relatively high despite our limited integration of weather data into the pipeline. This data could be collected at a more granular level provided an appropriate data service was discovered.

More experimentation is required to fully iterate through our features and the algorithms that are available and appropriate for the task. Further tweaking may yield better results. For example, we could sacrifice explainability in the case of an XGBoost or similar model being experimented with in the name of achieving stronger results. From an applied standpoint, concerns about transparency in the model's decision making would then lead to a requirement for further investigation over whether a black box model could feasibly replace the models chosen in this paper.

\subsection{Deployment and Further Evaluation}

Deployment of the project is already underway and is being carried out in a way that is conscious of Homeless Link's technical and financial capacity. The work done so far has led to various collaborations and the provision of resources to support the project in the mid to long-term, especially in overcoming some of the difficulties in ensuring that the models can work in real-time and be retrained at reasonable intervals. Homeless Link's status as a charitable organisation ensures eligibility for Microsoft's NGO Programme and so all of the work is to be hosted in an Azure environment and set up so that it can survive with a minimal requirement for expertise and maintenance.

Following deployment, it is necessary to ensure that the system has a positive impact in practice. This will take the form of a series of randomised control trials and observational studies to assess the impact that the work has on StreetLink's process over a year or more. These trials will be to test for significant changes in the positive outcome rate either directly, in terms of the alerts that are referred or not, and indirectly, by assessing whether the system shortens the time taken to make a decision and whether this has a positive impact on the outcomes of alerts following the assumption that the data will remain more relevant.

\subsection{Reproducibility}

The code, experiment definitions, and documentation are all available publicly at https://github.com/alan-turing-institute/DSSG19-HomelessLink-PUBLIC so that other organisations facing similar problems are able to use it as a starting point for similar projects. As previously stated, the data used by these models is sensitive in potentially identifying vulnerable individuals which means that trained models and the original historical data cannot be released. To try and mitigate the barrier this raises to furthering the research presented and to remain within the spirit of the conference, we have provided a synthetic dataset conforming to the shape and requirements of the code, as well as to the security and privacy-related requirements of Homeless Link. The broader problem of prioritisation and recommendation is common in social and outreach services, not just those tackling the issues of homelessness; we want other NGOs and government agencies to benefit from these findings through open sourcing this solution.

\section{CONCLUSIONS}

We present a valuable use case for machine learning in the area of social good via an approach to recommendation and risk classification comprised of two machine learning models. The Positive Outcome Model is for predicting whether an alert will result in a positive outcome within the next week, whilst the Referral Model is used for validation of the first and to ensure biases in the recommendations of the first are minimised or at least apparent. The model outperforms both of the defined baselines significantly and the current manual process at StreetLink by at least 15\% when compared to StreetLink's average person found rate for the past year. This translates to over 350 more rough sleepers being connected with per month during the busiest winter period. When smaller \textbf{\textit{k}} is considered, the performance improvement over the current manual process is significantly greater, suggesting that the top few received alerts could immediately be referred to alleviate some of Homeless Link's resource constraints. 

The Positive Outcome Model returns a ranked list of alerts allowing StreetLink staff to prioritise and augment their review process so as to more quickly and successfully deliver much needed aid to rough sleepers. Additionally, alerts sharing certain characteristics are shown to be currently underrepresented in the referrals made, but have predicted scores that suggest a positive outcome. Further work in trialling the solution will evaluate the impact of this prioritisation system on freeing up more resources and whether this exploration of currently underrepresented alerts leads to significant increases in the person found rate. Risk classification and prioritisation in this context allows organisations like Homeless Link to make better decisions on resource allocation, ensures StreetLink staff are dedicating their specialised training effectively, and maximises the overall positive impact that they can have on vulnerable people. The data-driven, evidential nature of our approach could also lead to positive policy change and longer-term impact.

\begin{acks}

To Gareth Thomas and Davide Veronese, both affiliated with Homeless Link, for their impassioned involvement and assistance throughout the project. Without the support of the organisation and these individuals this research would not be as close to deployment as it is now with the potential to positively impact thousands of lives in the UK.

All of this work was facilitated through the Data Science for Social Good Foundation's Fellowship programme running at The Alan Turing Institute and the University of Warwick in the summer of 2019. We owe our sanity to the 15 other fellows who inspired, supported and accompanied our team to the pub and beyond on numerous occasions, something that was often much needed. Many long nights and breakthroughs are owed to our mentors Adolfo De Unanue and Pablo Rosado, our project managers Joshua Sidgwick and Andrea Sipka, the organiser of the Warwick / Turing programme Sebastian Vollmer and to the founder of DSSG Rayid Ghani.

\end{acks}

\bibliographystyle{ACM-Reference-Format}
\bibliography{citations}


\begin{thebibliography}{16}


\ifx \showCODEN    \undefined \def \showCODEN     #1{\unskip}     \fi
\ifx \showDOI      \undefined \def \showDOI       #1{#1}\fi
\ifx \showISBNx    \undefined \def \showISBNx     #1{\unskip}     \fi
\ifx \showISBNxiii \undefined \def \showISBNxiii  #1{\unskip}     \fi
\ifx \showISSN     \undefined \def \showISSN      #1{\unskip}     \fi
\ifx \showLCCN     \undefined \def \showLCCN      #1{\unskip}     \fi
\ifx \shownote     \undefined \def \shownote      #1{#1}          \fi
\ifx \showarticletitle \undefined \def \showarticletitle #1{#1}   \fi
\ifx \showURL      \undefined \def \showURL       {\relax}        \fi
\providecommand\bibfield[2]{#2}
\providecommand\bibinfo[2]{#2}
\providecommand\natexlab[1]{#1}
\providecommand\showeprint[2][]{arXiv:#2}

\bibitem[\protect\citeauthoryear{{Arenas}, {Atkins}, {Austin}, {Beavan},
  {Cabrejas Egea}, {Carlysle-Davies}, {Carter}, {Clarke}, {Cunningham}, {Doel},
  {Forrest}, {Gabasova}, {Geddes}, {Hetherington}, {Jersakova}, {Kiraly},
  {Lawrence}, {Manser}, {O'Reilly}, {Robinson}, {Sherwood-Taylor}, {Tierney},
  {Vallejos}, {Vollmer}, and {Whitaker}}{{Arenas} et~al\mbox{.}}{2019}]%
        {2019TuringSafehaven}
\bibfield{author}{\bibinfo{person}{Diego {Arenas}}, \bibinfo{person}{Jon
  {Atkins}}, \bibinfo{person}{Claire {Austin}}, \bibinfo{person}{David
  {Beavan}}, \bibinfo{person}{Alvaro {Cabrejas Egea}}, \bibinfo{person}{Steven
  {Carlysle-Davies}}, \bibinfo{person}{Ian {Carter}}, \bibinfo{person}{Rob
  {Clarke}}, \bibinfo{person}{James {Cunningham}}, \bibinfo{person}{Tom
  {Doel}}, \bibinfo{person}{Oliver {Forrest}}, \bibinfo{person}{Evelina
  {Gabasova}}, \bibinfo{person}{James {Geddes}}, \bibinfo{person}{James
  {Hetherington}}, \bibinfo{person}{Radka {Jersakova}}, \bibinfo{person}{Franz
  {Kiraly}}, \bibinfo{person}{Catherine {Lawrence}}, \bibinfo{person}{Jules
  {Manser}}, \bibinfo{person}{Martin~T. {O'Reilly}}, \bibinfo{person}{James
  {Robinson}}, \bibinfo{person}{Helen {Sherwood-Taylor}},
  \bibinfo{person}{Serena {Tierney}}, \bibinfo{person}{Catalina~A. {Vallejos}},
  \bibinfo{person}{Sebastian {Vollmer}}, {and} \bibinfo{person}{Kirstie
  {Whitaker}}.} \bibinfo{year}{2019}\natexlab{}.
\newblock \showarticletitle{{Design choices for productive, secure,
  data-intensive research at scale in the cloud}}.
\newblock \bibinfo{journal}{\emph{arXiv e-prints}}, Article
  \bibinfo{articleno}{arXiv:1908.08737} (\bibinfo{date}{Aug}
  \bibinfo{year}{2019}), \bibinfo{numpages}{arXiv:1908.08737}~pages.
\newblock
\showeprint[arxiv]{cs.CR/1908.08737}


\bibitem[\protect\citeauthoryear{Blei, Ng, and Jordan}{Blei
  et~al\mbox{.}}{2003}]%
        {blei2003latent}
\bibfield{author}{\bibinfo{person}{David~M Blei}, \bibinfo{person}{Andrew~Y
  Ng}, {and} \bibinfo{person}{Michael~I Jordan}.}
  \bibinfo{year}{2003}\natexlab{}.
\newblock \showarticletitle{Latent dirichlet allocation}.
\newblock \bibinfo{journal}{\emph{Journal of machine Learning research}}
  \bibinfo{volume}{3}, \bibinfo{number}{Jan} (\bibinfo{year}{2003}),
  \bibinfo{pages}{993--1022}.
\newblock


\bibitem[\protect\citeauthoryear{Fetzer, Sen, and Souza}{Fetzer
  et~al\mbox{.}}{2019}]%
        {fetzer2019}
\bibfield{author}{\bibinfo{person}{Thiemo Fetzer}, \bibinfo{person}{Srinjoy
  Sen}, {and} \bibinfo{person}{Pedro~CL Souza}.}
  \bibinfo{year}{2019}\natexlab{}.
\newblock \showarticletitle{Housing insecurity, homelessness and populism:
  {Evidence} from the {UK}}.
\newblock  (\bibinfo{year}{2019}), \bibinfo{pages}{59}.
\newblock


\bibitem[\protect\citeauthoryear{Honnibal and Johnson}{Honnibal and
  Johnson}{2015}]%
        {honnibal2015improved}
\bibfield{author}{\bibinfo{person}{Matthew Honnibal} {and}
  \bibinfo{person}{Mark Johnson}.} \bibinfo{year}{2015}\natexlab{}.
\newblock \showarticletitle{An improved non-monotonic transition system for
  dependency parsing}. In \bibinfo{booktitle}{\emph{Proceedings of the 2015
  Conference on Empirical Methods in Natural Language Processing}}.
  \bibinfo{pages}{1373--1378}.
\newblock


\bibitem[\protect\citeauthoryear{{Institute of Global Homelessness}}{{Institute
  of Global Homelessness}}{2019}]%
        {igh2019}
\bibfield{author}{\bibinfo{person}{{Institute of Global Homelessness}}.}
  \bibinfo{year}{2019}\natexlab{}.
\newblock \bibinfo{booktitle}{\emph{State of {Homelessness} in {Countries} with
  {Developed} {Economies}}}.
\newblock
\urldef\tempurl%
\url{https://www.un.org/development/desa/dspd/wp-content/uploads/sites/22/2019/05/CASEY_Louise_Paper.pdf}
\showURL{%
Retrieved February 7, 2020 from \tempurl}


\bibitem[\protect\citeauthoryear{Meinbresse, Brinkley-Rubinstein, Grassette,
  Benson, Hall, Hamilton, Malott, and Jenkins}{Meinbresse
  et~al\mbox{.}}{2014}]%
        {meinbresse2014}
\bibfield{author}{\bibinfo{person}{Molly Meinbresse}, \bibinfo{person}{Lauren
  Brinkley-Rubinstein}, \bibinfo{person}{Amy Grassette},
  \bibinfo{person}{Joseph Benson}, \bibinfo{person}{Carol Hall},
  \bibinfo{person}{Reginald Hamilton}, \bibinfo{person}{Marianne Malott}, {and}
  \bibinfo{person}{Darlene Jenkins}.} \bibinfo{year}{2014}\natexlab{}.
\newblock \showarticletitle{Exploring the experiences of violence among
  individuals who are homeless using a consumer-led approach}.
\newblock \bibinfo{journal}{\emph{Violence and Victims}} \bibinfo{volume}{29},
  \bibinfo{number}{3} (\bibinfo{year}{2014}), \bibinfo{pages}{122--136}.
\newblock
\urldef\tempurl%
\url{https://doi.org/10.1891/0886-6708.vv-d-12-00069}
\showDOI{\tempurl}


\bibitem[\protect\citeauthoryear{{Ministry of Housing, Communities \& Local
  Government}}{{Ministry of Housing, Communities \& Local Government}}{2019}]%
        {govdefn}
\bibfield{author}{\bibinfo{person}{{Ministry of Housing, Communities \& Local
  Government}}.} \bibinfo{year}{2019}\natexlab{}.
\newblock \bibinfo{booktitle}{\emph{Governmental {Definitions} and {Notes} on
  {Homelessness}}}.
\newblock
\urldef\tempurl%
\url{https://www.gov.uk/guidance/homelessness-data-notes-and-definitions}
\showURL{%
Retrieved February 7, 2020 from \tempurl}


\bibitem[\protect\citeauthoryear{{Office for National Statistics}}{{Office for
  National Statistics}}{2016}]%
        {onsriskmanagement}
\bibfield{author}{\bibinfo{person}{{Office for National Statistics}}.}
  \bibinfo{year}{2016}\natexlab{}.
\newblock \bibinfo{booktitle}{\emph{{Working Paper 3: Risk Management}}}.
\newblock
\urldef\tempurl%
\url{https://www.ons.gov.uk/methodology/methodologytopicsandstatisticalconcepts/disclosurecontrol/healthstatistics}
\showURL{%
\tempurl}


\bibitem[\protect\citeauthoryear{Ortiz-Ospina and Roser}{Ortiz-Ospina and
  Roser}{2019}]%
        {ortiz-ospina_homelessness_2019}
\bibfield{author}{\bibinfo{person}{Esteban Ortiz-Ospina} {and}
  \bibinfo{person}{Max Roser}.} \bibinfo{year}{2019}\natexlab{}.
\newblock \bibinfo{booktitle}{\emph{Homelessness}}.
\newblock
\urldef\tempurl%
\url{https://ourworldindata.org/homelessness}
\showURL{%
Retrieved December 23, 2019 from \tempurl}


\bibitem[\protect\citeauthoryear{Pleace and Bretherton}{Pleace and
  Bretherton}{2018}]%
        {pleace_women_2018}
\bibfield{author}{\bibinfo{person}{Nicholas Pleace} {and}
  \bibinfo{person}{Joanne Bretherton}.} \bibinfo{year}{2018}\natexlab{}.
\newblock \showarticletitle{Women and {Rough} {Sleeping}: {A} {Critical}
  {Review} of {Current} {Research} and {Methodology}}.
\newblock  (\bibinfo{year}{2018}), \bibinfo{pages}{38}.
\newblock


\bibitem[\protect\citeauthoryear{{\v R}eh{\r u}{\v r}ek and Sojka}{{\v R}eh{\r
  u}{\v r}ek and Sojka}{2010}]%
        {rehurek_lrec}
\bibfield{author}{\bibinfo{person}{Radim {\v R}eh{\r u}{\v r}ek} {and}
  \bibinfo{person}{Petr Sojka}.} \bibinfo{year}{2010}\natexlab{}.
\newblock \showarticletitle{{Software Framework for Topic Modelling with Large
  Corpora}}. In \bibinfo{booktitle}{\emph{{Proceedings of the LREC 2010
  Workshop on New Challenges for NLP Frameworks}}}. \bibinfo{publisher}{ELRA},
  \bibinfo{address}{Valletta, Malta}, \bibinfo{pages}{45--50}.
\newblock


\bibitem[\protect\citeauthoryear{Roberts, Bahn, Ciuti, Boyce, Elith,
  Guillera-Arroita, Hauenstein, Lahoz-Monfort, Schr{\"o}der, Thuiller,
  et~al\mbox{.}}{Roberts et~al\mbox{.}}{2017}]%
        {roberts2017cross}
\bibfield{author}{\bibinfo{person}{David~R Roberts}, \bibinfo{person}{Volker
  Bahn}, \bibinfo{person}{Simone Ciuti}, \bibinfo{person}{Mark~S Boyce},
  \bibinfo{person}{Jane Elith}, \bibinfo{person}{Gurutzeta Guillera-Arroita},
  \bibinfo{person}{Severin Hauenstein}, \bibinfo{person}{Jos{\'e}~J
  Lahoz-Monfort}, \bibinfo{person}{Boris Schr{\"o}der},
  \bibinfo{person}{Wilfried Thuiller}, {et~al\mbox{.}}}
  \bibinfo{year}{2017}\natexlab{}.
\newblock \showarticletitle{Cross-validation strategies for data with temporal,
  spatial, hierarchical, or phylogenetic structure}.
\newblock \bibinfo{journal}{\emph{Ecography}} \bibinfo{volume}{40},
  \bibinfo{number}{8} (\bibinfo{year}{2017}), \bibinfo{pages}{913--929}.
\newblock


\bibitem[\protect\citeauthoryear{Toro}{Toro}{2007}]%
        {toro_toward_2007}
\bibfield{author}{\bibinfo{person}{Paul~A. Toro}.}
  \bibinfo{year}{2007}\natexlab{}.
\newblock \showarticletitle{Toward an {International} {Understanding} of
  {Homelessness}}.
\newblock \bibinfo{journal}{\emph{Journal of Social Issues}}
  \bibinfo{volume}{63}, \bibinfo{number}{3} (\bibinfo{date}{Sept.}
  \bibinfo{year}{2007}), \bibinfo{pages}{461--481}.
\newblock
\showISSN{0022-4537, 1540-4560}
\urldef\tempurl%
\url{https://doi.org/10.1111/j.1540-4560.2007.00519.x}
\showDOI{\tempurl}


\bibitem[\protect\citeauthoryear{Toro, Tompsett, Lombardo, Philippot,
  Nachtergael, Galand, Schlienz, Stammel, Yabar, Blume, MacKay, and
  Harvey}{Toro et~al\mbox{.}}{2007}]%
        {toro_homelessness_2007}
\bibfield{author}{\bibinfo{person}{Paul~A. Toro}, \bibinfo{person}{Carolyn~J.
  Tompsett}, \bibinfo{person}{Sylvie Lombardo}, \bibinfo{person}{Pierre
  Philippot}, \bibinfo{person}{Hilde Nachtergael}, \bibinfo{person}{Benoit
  Galand}, \bibinfo{person}{Natascha Schlienz}, \bibinfo{person}{Nadine
  Stammel}, \bibinfo{person}{Yanélia Yabar}, \bibinfo{person}{Marc Blume},
  \bibinfo{person}{Linda MacKay}, {and} \bibinfo{person}{Kate Harvey}.}
  \bibinfo{year}{2007}\natexlab{}.
\newblock \showarticletitle{Homelessness in {Europe} and the {United} {States}:
  {A} {Comparison} of {Prevalence} and {Public} {Opinion}}.
\newblock \bibinfo{journal}{\emph{Journal of Social Issues}}
  \bibinfo{volume}{63}, \bibinfo{number}{3} (\bibinfo{date}{Sept.}
  \bibinfo{year}{2007}), \bibinfo{pages}{505--524}.
\newblock
\showISSN{0022-4537, 1540-4560}
\urldef\tempurl%
\url{https://doi.org/10.1111/j.1540-4560.2007.00521.x}
\showDOI{\tempurl}


\bibitem[\protect\citeauthoryear{Varma and Simon}{Varma and Simon}{2006}]%
        {varma_bias_2006}
\bibfield{author}{\bibinfo{person}{Sudhir Varma} {and} \bibinfo{person}{Richard
  Simon}.} \bibinfo{year}{2006}\natexlab{}.
\newblock \showarticletitle{Bias in error estimation when using
  cross-validation for model selection}.
\newblock \bibinfo{journal}{\emph{BMC Bioinformatics}} \bibinfo{volume}{7},
  \bibinfo{number}{1} (\bibinfo{year}{2006}), \bibinfo{pages}{91}.
\newblock
\showISSN{14712105}
\urldef\tempurl%
\url{https://doi.org/10.1186/1471-2105-7-91}
\showDOI{\tempurl}


\bibitem[\protect\citeauthoryear{White and Maguire}{White and Maguire}{2018}]%
        {rsstats2019}
\bibfield{author}{\bibinfo{person}{John White} {and} \bibinfo{person}{Eva
  Maguire}.} \bibinfo{year}{2018}\natexlab{}.
\newblock \bibinfo{booktitle}{\emph{Rough Sleeping Statistics Autumn 2018,
  England (Revised)}}.
\newblock
\urldef\tempurl%
\url{https://www.gov.uk/government/statistics/rough-sleeping-in-england-autumn-2018}
\showURL{%
Retrieved January 21, 2020 from \tempurl}


\end{thebibliography}

\newpage

\appendix

\section{EXPERIMENTAL DETAILS}

The following supplementary material details what is required to reproduce our results as closely as possible. Note again that for legal and privacy reasons we can only release a privatised version of the dataset (scrubbed free text fields due to the possibility of personally identifiable information being present, and with anonymised outcomes) meaning results may vary. The experiments used in running the pipeline to generate our results are defined in YAML files that are also provided on the \href{https://github.com/alan-turing-institute/DSSG19-HomelessLink-PUBLIC}{aforementioned public GitHub repository}. The repository also includes a more in-depth description of running the pipeline and the requirements of a system to do so.

\subsection{MODEL TRAINING}

Grid searches of the following pairings of parameter spaces and Scikit-Learn implementations of algorithms were carried out:

\begin{itemize}
    \item Random Forest, Extra Trees using the Gini Impurity criterion and setting the maximum number of features to be the square root of the total: \begin{itemize}
        \item Number of Estimators: 100, 250, 500, 1000, 2500, 5000, 7500, 10000
        \item Maximum Tree Depth: 1, 2, 3, 4, 5, 10 and None
    \end{itemize}
    \item AdaBoost: \begin{itemize}
        \item Number of Estimators: 100, 250, 500, 1000, 2500, 5000, 7500, 10000
        \item Learning Rate: 0.01, 0.05, 0.1, 0.25, 0.5, 1.0
    \end{itemize}
    \item Decision Trees: \begin{itemize}
        \item Max Depth: 1, 2, 3, 4, 5, 6, 7, 8, 9 and None
    \end{itemize}
\end{itemize}

\subsection{Cross Validation Configuration}

Month forward-chaining temporal cross validation was used in all of our final experiments with the following parameters:

\begin{itemize}
    \item Data Start Point: 2017-12-01
    \item Data End Point: 2019-02-28
    \item Training Data Label Span: 1 Week
    \item Test Data Label Span: 1 Week
    \item Training Data Span: 2 Years
    \item Training Frequency: 1 Day
    \item Test Data Span: 1 Month
    \item Testing Frequency: 1 Day
    \item Model Update Frequency: 1 Month
\end{itemize}

\subsection{FEATURES}

All of the generated and extracted features were used in the final experiments in order to also get a full view of feature importance in our final models.

\end{document}